
\input phyzzx
\input epsf
\overfullrule=0pt
\hsize=6.5truein
\vsize=9.0truein
\voffset=-0.1truein
\hoffset=-0.1truein

%
%
\def\half{{1\over 2}}
\def\IC{{\ \hbox{{\rm I}\kern-.6em\hbox{\bf C}}}} \def\IR{{\hbox{{\rm
I}\kern-.2em\hbox{\rm R}}}} \def\IZ{{\hbox{{\rm  
Z}\kern-.4em\hbox{\rm Z}}}}

\def\sIR{{\hbox{{\sevenrm I}\kern-.2em\hbox{\sevenrm R}}}}

\def\n.{{N \over 2}}

\def\rw{Robertson-Walker}
\def\hp{holographic principle}
\def\es{equation of state}
\def\chisubH{\chi\lower.5ex\hbox{$_H$}}

%
%
\hyphenation{Min-kow-ski}
\rightline{SU-ITP-98-39,UTTG-06-98}
\rightline{hep-th/9806039}
\rightline{June 1998}

\vfill

%
%
\title{ Holography and Cosmology}

\vfill

%
%
\author{W. Fischler$^1$}

\vfill

\address{$^1$Theory Group,Department of Physics,University of Texas\break
Austin,TX 78712}
\vfill

\author{L. Susskind$^2$}

\vfill

\address{$^2$Department of Physics,
Stanford University\break Stanford, CA
94305-4060}
\vfill


\vfill

\vfill

%
%
A cosmological version of the holographic principle is proposed. Various
consequences
are discussed including bounds on equation of state and the requirement
that the universe be infinite.


%
%



\REF\holo{L. Susskind, The World as a Hologram, hep-th/9409089}


\REF\hoof{ C.R. Stephens, G.
't Hooft and B.F. Whiting, Black Hole Evaporation Without  
Information Loss,
gr-qc/9310006}

\REF\witt{ Edward Witten, Anti De Sitter Space And Holography,
hep-th/9802150 }
\REF\ads{L. Susskind and E. Witten, The Holographic Bound in Anti-de Sitter Space, hep-th/9805114}
\REF\kas{ E.Kasner, Am.J.Math 48,217 (1921) }
\REF\chod{ A.Chodos and S.Detweiler, Phys.Rev.D21,2167 (1980) }


%
%

%
%
\chapter{Introduction}

The recent revolution coming from string theory and black hole theory has
taught us many
unexpected things about the nature of spacetime and its relation to  
matter,
energy and
entropy. Such a conceptual paradigm shift must eventually have serious
implications for cosmology.

Perhaps, the most radical modification of standard concepts  
required by the
new theory is the \hp \  [\holo] [\hoof] [\witt] [\ads]. The \hp \ requires
that the degrees of
freedom of a spatial region reside not in the interior as in an ordinary
quantum field theory
but on the surface of the region. Furthermore it requires the number of
degrees of freedom per
unit area to be no greater than 1 per Planck area. As a consequence, the
entropy of a region
must not exceed its area in Planck units. Thus far the \hp \ has  
not found
its way into
cosmology. In this paper we will try to correct this situation with some
preliminary thoughts on
how to formulate the \hp \ in cosmology and discuss some of its  
consequences.
\chapter{Flat Universe}

For most of this paper it will be assumed that spacetime is described by
the usual \rw \
cosmology with space being flat. The metric has the usual form. $$
ds^2 = dt^2 - a^2(t) dx^idx^i
\eqn\one
$$
The number of spatial dimensions will be kept general so that $ i$ runs
from $1$ to $d$.
The cosmological assumptions are the ususal ones including homogeneity,
isotropy and in the late
universe constant entropy density in comoving coordinates.

Let us begin with a proposal that we will very quickly rule out; The
entropy in any region of
coordinate size $ \Delta x \sim R$ never exceeds the area. Since our
assumptions require the
entropy density to be constant, the entropy in a region is  
proportional to
its coordinate volume
$R^d$. Furthermore in flat space the surface area of the region  
grows with
$R$ like
$[Ra(t)]^{d-1}$. Obviously when $R$ becomes large enough, the entropy
exceeds the area and the
principle is violated.

A more sophisticated version goes as follows. Consider a spherical spatial
region $\Gamma$ of
coordinate size $R$ with boundary $B$. Now consider the light-like  
surface
$L$ formed by past
light rays from $B$ toward the center of $\Gamma$. There are three
situations. If $R$ is smaller
than the coordinate distance to the cosmological horizon $R_H$ then the
surface $L$ forms a
light cone with its tip in the future of the singularity at $t=0$. If $R$
is the size of the
horizon then $L$ is still a cone but with its tip at $t=0$. However if
$R>R_H$ the surface is
a truncated cone. Now consider all the entropy (particles) which pass
through $L$. For the
first two cases this is the same as the entropy in the interior of  
$\Gamma$
at the instant $t$.
But in the last case the entropy within $\Gamma$ exceeds the entropy
passing through $L$. The
proposed \hp \ is that the entropy passing through $L$ never exceeds the
area of the bounding
surface $B$. It is not difficult to see that for the homogeneous  
case, this
reduces to a single
condition:

The entropy contained within a volume of coordinate size $R_H$ should not
exceed the area of
the horizon in Planck units. In terms of the (constant) comoving  
entropy density $\sigma$ $$
\sigma R_H^d <[ aR_H]^{d-1}
\eqn\two
$$
with every thing measured in Planck units. Note that both $R_H$ and  
$a$ are
functions of time and
that
\two
\ must be true at all time.

Let us first determine whether \two \ is true today. The entropy of the
observable universe is
of order $10^{86}$ and the horizon size (age) of the universe is of order
$10^{60}$. Therefore
the ratio of entropy to area is much smaller than 1. Now consider whether
it will continue to be
true in the future. Assume that $a(t) \sim t^p$. The horizon size is
determined by
$$
R_H(t) = \int_0^t {dt \over a(t)} \sim t^{1-p} \eqn\three
$$
Thus in order for \two \ to continue to be true into the remote future we
must satisfy
$$
p>{1 \over d}
\eqn\four
$$
In other words there is a lower bound on the expansion rate.

The bound on the expansion rate is easily translated to a bound on the
equation of state. Assume
that the \es \ has the usual form
$$
P = \gamma \epsilon
\eqn\five
$$
where $P$ and $\epsilon$ are pressure and energy density. Standard  
methods
yield a solution of the
Einstein equations
$$
a(t) \sim t^{2 \over d(1+ \gamma)}
\eqn\six
$$
Thus $p ={2 \over d(1+ \gamma)}$ and the inequality \four \ becomes $$
\gamma<1
\eqn\seven
$$
The bound \seven \ is well known and follows from entirely different
considerations. It
describes the most incompressible fluid that is consistent with special
relativity. A violation
of the bound would mean that the velocity of sound exceeds the  
velocity of
light. We will take
this agreement as evidence that our formulation of the cosmological \hp \
is on the right track.

Although violating the bound  \seven \ is impossible, saturating it  
is easy.  We will give two examples.  In the first example the  
energy density of the universe is dominated by a homogeneous   
minimally coupled scalar field. It is well known that in this case  
the pressure and energy density are equal.

Another example involves  flat but anisotropic
universes,for example the
anisotropic (Kasner [\kas] [\chod]) universes with metric:
$$
ds^2 = dt^2 - \Sigma_i\ {t^{2p_i} {dx_i}^2}
\eqn\eight
$$
The ratio entropy/area, $S/A$ ,for this case is easily evaluated to be:
$$
S/A = \Pi_i \ R_{H,i}~ / {[\Pi_j\ {t^{p_j}t^{1-p_j}]}^{{d-1}/d}}
\eqn\nine
$$
where $R_{H,i} = t^{1-p_i}$ is the coordinate size of the horizon in
direction i . The denominator
in equation \nine \ is the proper area of the horizon.

Equation \nine \ gives:
$$
 S/A = t~ ^ {1 - \Sigma_i\ p_i }
\eqn\ten
$$

 The conditions on the exponents of the Kasner solutions are
obtained by using the Einstein equations.The
exponents  satisfy the
following equations:
$$
\Sigma_i\ p_i = 1
 \eqn\eleven
$$
$$
 \Sigma_i\ p_i^2 = 1
 \eqn\twelve
$$
it then follows from \eleven \ and \twelve \ that $ S/A $ for these flat
anisotropic universes is
constant in time.Thus depending on the boundary condition the \hp \  
may be saturated by these universes.

Having established that the \hp \ will not be violated in the future we
turn to the past. First
consider the entropy-area ratio ($\rho = S/A$) at the time of decoupling.
Standard estimates give
a ratio which is about $10^6$ times bigger than todays ratio. Thus at
decoupling $$
\rho(t_d) \sim 10^{-28}
\eqn\thirteen
$$
During a radiation dominated era it is easily seen that $\rho$ is
proportional to $t^{-\half}$.
$$
\rho = 10^{-28} \left[{t_d \over t}\right]^\half
\eqn\fourteen
$$
Remarkably, in Planck units $t_d^{\half} \sim 10^{28}$ so that $\rho < 1$
for all times later
than the planck time. The entropy in the universe is as large as it  
can be
without the \hp \
having been violated in the early universe!
\chapter{Non Flat Universe}

Let us now turn to non flat universes and consider first the case of a
closed universe where for
simplicity we will restrict the discussion to 3 + 1 dimensions. The  
metric
has the form
$$
ds^2 = dt^2 - a^2(t)(d\chi^2 + \sin^2\chi~ d\Omega^2)
\eqn\fifteen
$$
where  $\chi$ and $\Omega$ parametrize the three-sphere, $S_3$ .  The
azimuthal angle of
$S_3$ is denoted by $\chi$ and $\Omega$ is the solid angle parametrizing
the two-sphere at fixed
$\chi$.
 We can then proceed in calculating the entropy/area , $S/A$ ,by  
following
the procedure outlined in
the flat case.
$$
 S/A = {2\chisubH - \sin 2\chisubH \over 2a^2(\chisubH)\sin^2(\chisubH)}
\eqn\sixteen
$$
where  $\chisubH = \int dt/a(t)$    is the coordinate size of the  
horizon.

As the universe evolves
it  will inevitably  reach a stage when it will saturate and then threaten to
violate the
holographic bound .
This can be easily seen by considering  an equation of state such  
that the
energy density scales
like $ a^{-2K} $,with $K>1$.
Solving the Hubble equation gives
$$
{a^{K-1}(\chisubH)} \sim \sin (K-1)\chisubH
\eqn\seventeen
$$
The holographic bound $S/A < 1$ implies:
$$
a^2(\chisubH)\geq {2\chisubH - \sin 2\chisubH\over 2 \sin^2(\chisubH)}
\eqn\eightteen
$$
So as $\chisubH$ approaches  $\pi/(K-1)$ , the two previous equations \seventeen\
and \eightteen\ cannot be
consistent.Depending on the equation of state, the bound will be reached either while the universe
is still growing ,for example when the energy density is dominated by non-relativistic matter
 or during recollapse like in a radiation dominated universe . This
seems to indicate that positively curved closed   universes are inconsistent with the \hp \ . We do
not know what new behaviour sets in to accomodate   the \hp
\ or if this violation
of the \hp \
 just excludes these universes as inconsistent?

The case of negatively curved open universes can be studied in a similar
manner. At early times,
both open and closed universes look flat as long as the energy density
scales like $a^{-2K}$ and $K
> 1$ .During that period the holographic bound implies that the speed of
sound $< c$ . At
later times the holographic bound for open universes does not put a  
strong
bound on the equation of
state. Since for "late" times the area and the volume grow in fixed
proportion ,there is then no
bound on how slow the expansion has to be.

 \chapter{Acknowledgments}

W.F. was supported in part by the Robert Welch Foundation and NSF  
Grant PHY
- 9511632. L.S.
acknowledges the support of the NSF under Grant No. PHY - 9219345.

\refout
\end